\documentclass[preprintnumbers,amsmath,latexsym,array,enumerate,letter,11pt]{revtex4}
\usepackage{subfigure}
\usepackage{graphicx}
\usepackage{dcolumn}
\usepackage{bm}
\usepackage{amssymb}
\usepackage{epsfig}
\usepackage{slashed}
\usepackage[utf8]{inputenc}

\newcommand{\half}{\frac{1}{2}}
\newcommand{\be}{\begin{equation}}
\newcommand{\ee}{\end{equation}}
\newcommand{\beq}{\begin{eqnarray}}
\newcommand{\eeq}{\end{eqnarray}}

\newcommand{\mybar}[1]%
        {\kern  .6pt\overline{\kern - .6pt#1\kern - .6pt}\kern  .6pt}
\begin{document}
\title{Large Two-loop Effects in the Higgs Sector as New Physics Probes}
\author{Sichun Sun}
\email{sichun@uw.edu}

\affiliation{Jockey Club Institute for Advanced Study, Hong Kong University of Science and Technology, Clear Water Bay, Hong Kong}

\date{\today}

\begin{abstract}
We consider a simple Higgs portal model in beyond the standard model scenario: an extra real gauge singlet scalar that couples to the Higgs.  We calculate the higher-loop corrections to the cross section of the Higgsstrahlung process $e^+e^- \rightarrow Z h$, along with the tri-Higgs coupling and the wave-function renormalization. We find noticeable contribution to the total Higgsstrahlung cross section, especially coming from two-loop diagrams. We also find that the correction to the tri-Higgs coupling is complex when this extra scalar is lighter than half of the center of mass energy. It indicates a new source of CP violation. In the region of this extra scalar being lighter than a couple of hundreds of GeVs, we argue that the higher-loop calculation is a more reliable approach than the effective field theory calculation.
\end{abstract}

\maketitle

\section{introduction}
Since the successful and exciting Higgs discovery \cite{Chatrchyan:2012xdj,Aad:2012tfa}, studies of Higgs properties have been the focus of high energy physics. The Higgs physics: the Higgs related couplings and processes now have become our major guide to new physics beyond the standard model (BSM) and will be further explored in the next coming years.

Many couplings of the Higgs have already been constrained by the large hadronic collider (LHC), including the Higgs couplings to fermions, gluons and electroweak gauge bosons. All of them are within 20$\%$ of the SM prediction \cite{2015arXiv150704548A,Khachatryan:2014jba,Henning:2014gca,Dawson:2002wc,Peskin:2013xra,Robens:2015gla,2015EPJC...75..212K}. The impending Run-2 of the LHC, and the higher luminosity upgrade program will improve the current measurements and take additional measurements. Among them, the tri-Higgs coupling is an anticipated new type of interaction. Studies show in high luminosity LHC, $30 \%$ to $50 \%$ accuracy of the measurement can be achieved \cite {2013,2013JHEP...04..151B,2013JHEP...06..016G,2014PhLB..728..433B} and down to around $10 \%$ in the future hadronic collider designs \cite{2013arXiv1308.6302Y,2015JHEP...02..016B}.

However, due to the overwhelming hadronic processes, it is hard to obtain Higgs measurements in hadronic machines below a few percents both due to systematic and theoretical errors. Meanwhile the Higgs coupling measurements depend on Higgs decay rates at LHC which are model dependent, since the total width of the Higgs is not a direct observable. To measure the Higgs couplings with better precision, proposed future precision high-energy $e^+e^-$ machines, e.g, ILC, CEPC,and FCC-ee (TLEP) are needed. Experimentally, the events are very clean in those leptonic machines. The SM precision measurements can be obtained without hadronic processes. Moreover, the cross section of $e^+e^- \rightarrow Z h$ is proved to be a robust, model-independent measurement through Z tagging \cite{Hagiwara:2000tk}. Those facts make the cross section of $e^+e^- \rightarrow Z h$ a good observable to probe the physics beyond the SM. In table I we show a list of Higgsstrahlung constraints for different future $e^+e^-$ collider proposals.

There are extensive previous studies on anomalous Higgs couplings, and their impact on the Higgsstrahlung process \cite{Beneke:2014sba,Hagiwara:1993sw,Gounaris:1995mx,Kilian:1996wu,GonzalezGarcia:1999fq,Kile:2007ts,Fleischer:1982af,Kniehl:1991hk,Carena:1995bx,Aoki:1982ed,Pomarol:2013zra,Elias-Miro:2013eta,McCullough:2013rea,Craig:2013xia,Craig:2014una,Englert:2013tya}. A complete one loop SM calculation has been done more than 20 years ago by A. Denner \cite{Denner:1992bc,Denner:1991kt}. Unlike the fact that studies of QCD cross section have gone even beyond the NNLO, because of the perturbative nature of electroweak physics and lack of experimental measurements, people hardly go beyond one-loop in Higgsstrahlung process. Here we consider a well-studied BSM model, adding $H^\dagger H \Phi^2$ to the Lagrangian with $\Phi$ being a singlet scalar. This model belongs to the Higgs portal class and $\Phi$ could also couple to the dark matter \cite{Patt:2006fw,Burgess:2000yq,Englert:2011yb,Djouadi:2011aa,Chacko:2013lna,Greljo:2013wja,Cline:2013gha}. It also has the ability to change the electroweak symmetry breaking dynamics \cite{Grojean:2004xa,Katz:2014bha,Curtin:2014jma}, achieving a first-order phase transition. This simple model can also be embedded into many UV-completed models, including supersymmetry \cite{Cohen:1996vb,Dimopoulos:1995mi}, and stringy ones.

In this paper, we study this simple yet well-motivated SM extension, and go beyond one-loop effects. There has been previous attempt that went beyond one-loop level by using an effective field theory approach \cite{Barger:2003rs,Elias-Miro:2013mua,Falkowski:2014tna}. Those approaches are valid only when the new particles are heavier than a few hundred GeV\textemdash the scale that lepton colliders are probing. This is due to the nature of the effective theory: integrating out heavy particles/short distance. The lighter scalar region is experimentally valid in some cases, and interesting phenomenologically, e.g., serving as some dark matter portals as in \cite{Chacko:2013lna,Greljo:2013wja,Cline:2013gha}.

\begin{table}
\begin{tabular}{||c| c| c|c|c||c||c|| } 
Colliders&\multicolumn{4}{c||}{ILC}   \\
\hline
Scienarios & 250/fb @250GeV&500/fb @500 GeV&1.15/ab @250GeV&1.6/ab @500GeV \\
\hline
$\frac{\delta \sigma_{ZH}}{\sigma_{ZH}}$&$2.6\%$&$3.0\%$ &$1.2\%$ &$1.7\%$  \\
 \end{tabular}
 \begin{tabular}{||c|c|c|| } 
Colliders & CEPC &FCC-ee  \\
\hline
Scienarios&5/ab @240GeV & 10/ab @240GeV \\
\hline
$\frac{\delta \sigma_{ZH}}{\sigma_{ZH}}$ &$0.5\%$ &$0.4\%$ \\
 \end{tabular}
\caption{The constraints of Higgsstrahlung process in the proposed future electron colliders. The percentage data are from \cite{Asner:2013psa,Dawson:2013bba,Baer:2013cma} for ILC, \cite{CEPC:2015} for CEPC and \cite{Gomez-Ceballos:2013zzn} for FCC-ee.} 
\label{tab:template} 
\end{table}

The paper is organized as follow: in section II we lay down the theory framework of this model. We present the possible CP violating tri-Higgs coupling correction and the Higgs wavefunction correction to the SM couplings. In section III we present the calculation of the Higgsstrahlung process up to two-loop level in this BSM scenario. Those contributions can be divided into three groups of different diagrams, and part of those diagrams can be reduced to one-loop diagrams. Some special packages are used to evaluate the loop integrals, and the results are semi-numerical. In section IV we briefly discuss the impact on the electroweak baryogenesis. We conclude in section V.

\section{The theoretical framework}
In this section, we outline the general framework of our discussion, and present two useful results on tri-Higgs coupling and the analysis on the Higgs wavefunction renormalization for later use.  Especially we deal carefully with the small mass region of the extra scalar.

\subsection{The Higgs potential and the correction to the couplings}

In this work, we assume a single SM Higgs doublet H with a general renormalizable tree-level Higgs potential, responsible for electroweak symmetry breaking:
\beq
V_0= -\mu^2 H^2 +\lambda H^4
\eeq
Substituting $H=(H^+, (v+h+iA^0)/\sqrt{2})$, and going to the unitary gauge now, which sets $H^+, A^0$ to zero. Note that we only do this for the tree and one-loop related calculation. We switch to t'Hooft-Feynman gauge for the two-loop calculation later. In unitary gauge we have: 

\beq
&V_0=-\half m_h^2 h^2-\frac{ m_h^2}{2v}h^3-\frac{1}{8} (\frac{m_h}{ v})^2 h^4\\
& =-\half m_h^2 h^2-\frac{1}{3!}3 m_h^2v^{-1} h^3-\frac{1}{4!}3 m_h^2 v^{-2} h^4
\eeq

The Eq.3 above includes the symmetry factor, for the Feynman diagram calculation later. In the choice of our parametrization, the Higgs acquires a VEV $<h>=v=\mu/\sqrt{\lambda} \approx 246$ GeV. A tree level Higgs mass is $m_h= \sqrt{2}\mu\approx 125$GeV. Then the coefficients of the Higgs potential are $\lambda\approx 0.13$ and $\mu \approx 90$ GeV.

One includes an extra singlet scalar $\Phi$ and a $Z_2$ symmetry under which $\Phi\rightarrow -\Phi$. It couples to the Higgs field as below: 

\beq
L \supset -\kappa H^2\Phi^2-\half \mu_\Phi^2 \Phi^2
\eeq
After electroweak symmetry breaking, and taking the SM Higgs part, this becomes:
\beq
L\supset-\frac{1}{2!2!}2\kappa h^2\Phi^2-\half 2\kappa v h \Phi^2-\half (\mu_\Phi^2+\kappa v^2) \Phi^2
\eeq

We call this model \textquotedblleft$Z_2$ symmetric SM+S" scenario. It has been discussed in different aspects, to solve problems including "naturalness", electroweak symmetry breaking, baryogenesis and WIMP dark matter. This simple  model can arise from various BSM scenarios. In this work we adopt the most general and simple form of the model, and focus on the Higgs-related higher loop effects. The higher loop corrections get directly into the precision Higgsstrahlung process and potentially contribute to the EWBG with noticeable size. 

The relevant parameters in this model are $\mu_\Phi$ and $\kappa$. Here we assume that:
\beq
m_\Phi^2=\mu_\Phi^2+\kappa v^2 > 0
\eeq
From now on we treat $m_\Phi$ as the physical mass and deal with it rather than $\mu_\Phi$. We also focus on $m_\Phi > m_h/2$ region, since otherwise it is very constrained from Higgs decays branching ratio data. 

\subsection{The Correction to tri-Higgs coupling}
\begin{figure}[t]
\includegraphics[width=7cm,natwidth=610,natheight=632]{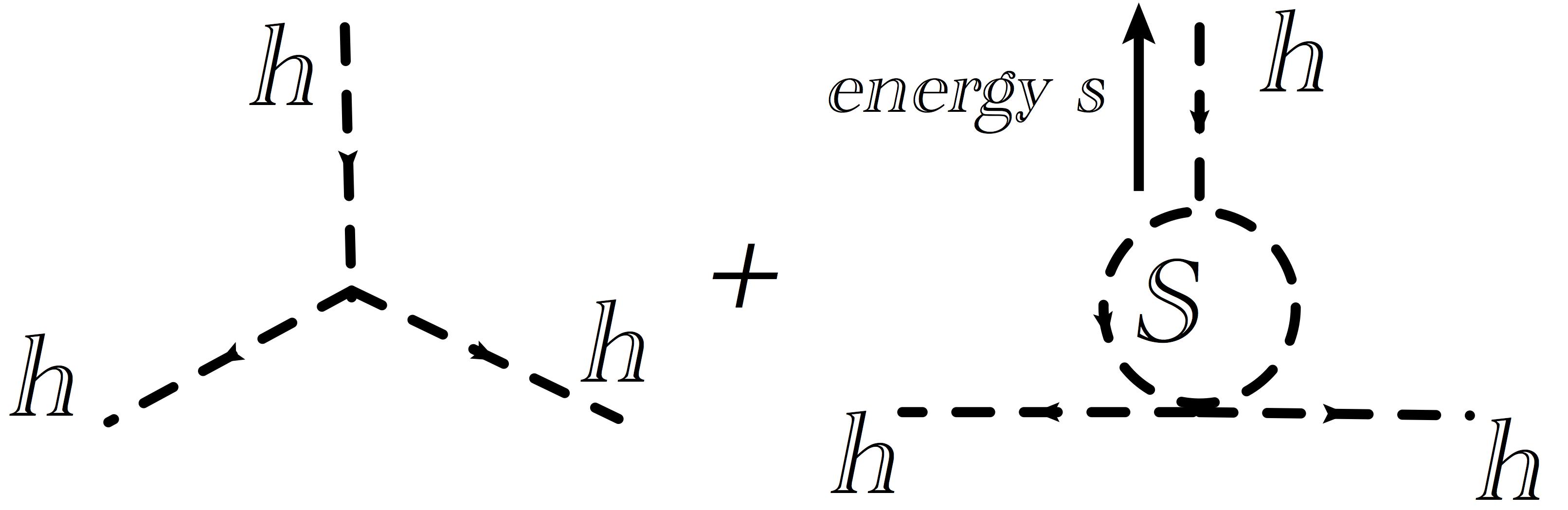}
\caption{The Feynman diagram of tri-Higgs coupling correction, from the extra scalar.}
\label{fig:tri1}
\end{figure}
\begin{figure}[t]
\includegraphics[width=9cm,natwidth=610,natheight=632]{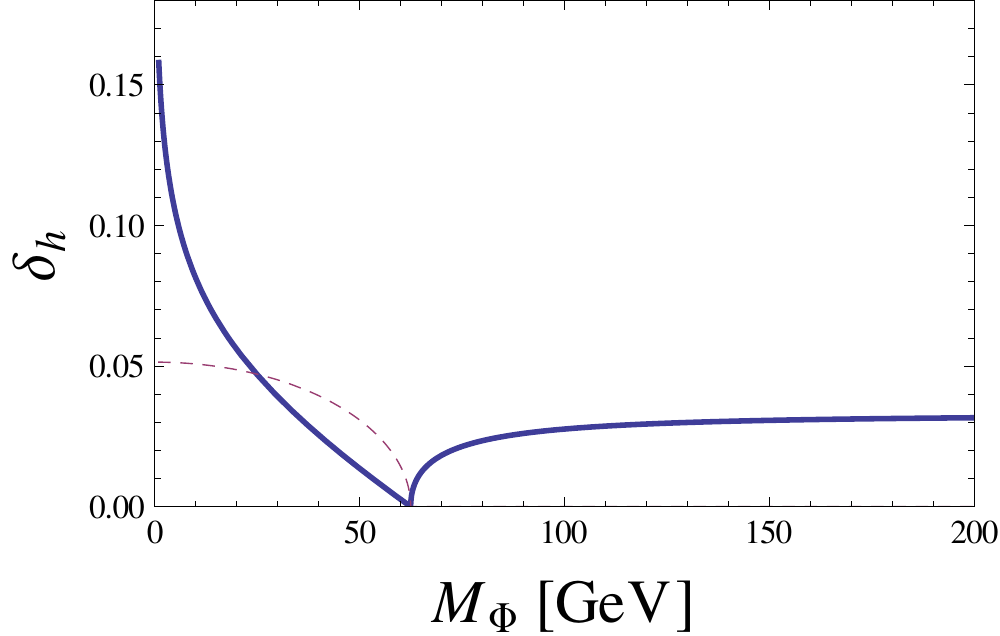}
\caption{The real (Thick line) and imaginary (Dashed line) part of $\delta_h$ versus $m_\Phi$ with $s=m_h^2$.}
\label{fig:tri2}
\end{figure}
We start with the one-loop effect induced by the extra scalar in this model. The tri-Higgs coupling has drawn much attention after the discovery of the Higgs boson \cite{2013,2013JHEP...04..151B,2013JHEP...06..016G,2014PhLB..728..433B}. People consider it as a possible new fundamental force: the self-interaction of the spin-0 particles. 

We show all the one-loop and two-loop Feynman diagrams in Fig.\ref{fig::twoloop}. One should notice that the correction to tri-Higgs coupling induced by the extra particle in this case is not a constant. Instead it depends on the 4-momentum of the Higgs in the $H\Phi \Phi$ vertex. Although it is sometimes good to get an estimate by integrating $\Phi$ out, we can not always do that unless $\Phi$ is much heavier than the energy of the process. However notice that the loop integral DOES reduce to a constant when the Higgs propagator  of the $H\Phi\Phi$ coupling is on-shell with one of the outgoing particle energy $s=m_h^2$, which happens in Graph c and d in Fig \ref{fig::twoloop}.

Here we define the vertex function with the tree level value and the loop correction as in Fig 1:
\beq
&V(s,m_\Phi,m_h)=-3\frac{m_h^2}{v}i+\half (-2\kappa i)(-2\kappa v i)\int \frac{d^4k}{(2\pi)^4}\frac{i}{k^2-m_\Phi^2}\frac{i}{(k+p)^2-m_\Phi^2}\\
&=-3\frac{m_h^2}{v}i- \frac{\kappa^2 v i}{8 \pi^2} \int^1_0 dx Log(\frac{m_\Phi^2-x(1-x)s}{m_\Phi^2-x(1-x)4m_\Phi^2})
\eeq

Defining a function $\delta_h(m_\Phi,s)$ as the correction to the SM tri-Higgs coupling as $ -(1+\delta_h \kappa^2) \frac{3m^2_h}{v} h^3 $, we have:
\beq
\delta_h(m_\Phi,s)=(V(s,m_\Phi,m_h)/V_{SM}-1)/\kappa^2=\frac{v^2}{m_h^2 24 \pi^2}\int^1_0 dx Log(\frac{m_\Phi^2-x(1-x)s}{m_\Phi^2-x(1-x)4m_\Phi^2})
\eeq

We plot the $\delta_h (s)$ as the function of $m_\Phi$ in Fig.\ref {fig:tri2}. We use this result to simplify some two-loop diagrams later. It also shows that with much lighter $m_\Phi$, the correction to tri-Higgs coupling increases dramatically.

\subsection{The wave function renormalization}

\begin{figure}[t]
\includegraphics[width=9cm,natwidth=610,natheight=632]{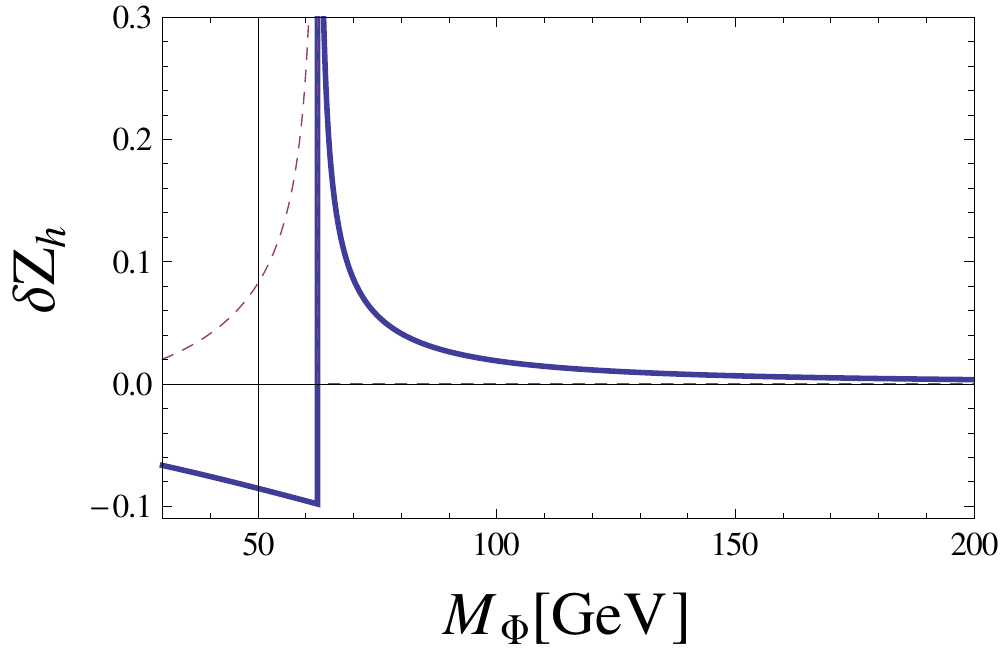}
\caption{The real (Thick line) and imaginary (Dashed line) part of $\delta Z_h$ versus $m_\Phi$ in GeVs with $\kappa=1$. Note that there are poles for both parts at $m_\Phi=m_h/2$. It is the resonance of the two extra scalar bound state. The imaginary part goes to zero when the branch of the Higgs decaying into two $m_\Phi$ closes within $m_\Phi> m_h/2$.}
\label{fig:tri}
\end{figure}

The extra scalar field $\Phi$ renormalizes the Higgs field at one loop level, and no other one-loop corrections with this scalar at this level. Then one can rescale $h \rightarrow (1-\delta Z_h/2)h$. It rescales all Higgs couplings, including cubic and quartic couplings, and all Higgs couplings to weak gauge bosons and fermions. This rescaling is physical. The corrections to one-loop processes due to $Z_h$ is considered to be two-loop effects by counting $1/(4\pi)^2$. These one-loop integrals can be generated and calculated in Feyncal, FeynArt and Looptools \cite{Hahn:2000kx,Hahn:1998yk}. We define:

\beq
\delta Z_H = - Re \frac{\partial \Sigma^H (k^2)}{\partial k^2}\mid_{k^2=M^2_H}
\eeq

With
\be 
i\Sigma= \half (-i 2 \kappa v)^2\int \frac{d^4k}{(2\pi)^4}\frac{i}{k^2-m_\Phi^2}\frac{i}{(k+p)^2-m_\Phi^2}=\half (-i 2 \kappa v)^2 B(p^2,m_\Phi^2,m_\Phi^2)\frac{i}{16\pi^2}\ee

\beq
\delta Z_H = -Re[\half (-i 2 \kappa v)^2\frac{1}{16\pi^2} \partial B(p^2,m_\Phi^2,m_\Phi^2)/\partial p^2\mid_{p^2=M^2_H}]
\eeq

Here the $B(p^2, m_\Phi^2,m_\Phi^2)$ is defined and evaluated in Feyncal. We plot the value of $\delta Z_h$ in Fig.3. At tree level, the HZZ coupling becomes $ieg_{\mu \nu} M_w\frac{1}{sc^2}(1+\half \delta Z_H)$. $\delta Z_H$ correction gets into all the Higgs related couplings in form of $\half \delta Z_H$ per Higgs line. We list all the vertex corrections in Table.\ref{tab:template}.

 \begin{table}
\begin{tabular}{c| c| c|c|c|} 
vertices & $h^4$ & $h^3$ & $ZZhh$ & $W^+W^-hh$ \\
\toprule
couplings & $-\frac{3ie^2}{4s^2}\frac{M_H^2}{M_W^2}(1+2\delta Z_H)$ &
$-\frac{3ie}{2s}\frac{M_H^2}{M_W}(1+\frac{3}{2}\delta Z_H)$&
$\frac{g_{\mu\nu}ie^2}{2s^2c^2}(1+\delta Z_h)$
&$\frac{ie^2g_{\mu\nu}}{2s^2}(1+\delta Z_h)$  
\end{tabular}
\begin{tabular}{c| c| c } 
vertices & $hZZ$ & $h\bar{f}f$\\
\toprule
couplings 
&$\frac{ie g_{\mu\nu}M_w}{sc^2}(1+\half \delta Z_h)$ 
& $\frac{-ie}{2s M_w}\delta _{ij} m_{f,i}(1+\half \delta Z_h)$ \\ 
\end{tabular}
\caption{The tree level SM vertices values with the correction from Higgs wave-function renormalization by the extra scalar assuming $\kappa=1$ in t'Hooft Feynman gauge. s and c are the sine and cosine function of the weak angle.} 
\label{tab:template} 
\end{table}

All the one-loop diagrams are calculated in \cite{Denner:1991kt}. It includes the correction from wave-function renormalization in the SM. We plot all the two loop effect combined in Section IV, after the discussion about other diagrams.

\section{Two-loop effect in Higgsstrahlung and beyond}

\subsection{Some nontrivial loop integrals}
We show all the diagrams up to two-loop level with $\Phi$ scalar in Fig.\ref{fig::twoloop}. We notice that some of those can be reduced to one-loop diagrams plus corrections: the value of diagram c and d can be evaluated by the diagram a and b with a constant tri-Higgs coupling correction. 

For Graph e,f,g, however, we can not reduce them to one-loop diagrams. Notice that at two-loop level when we are doing the calculation of Graph e, f, g in t'Hooft-Feynman gauge, there are some non-standard Higgs related vertices involving eaten Goldstone boson, e.g., $H^+H^- \Phi\Phi, A^0A^0\Phi\Phi$. However those vertices are not present in our case.  In t'Hooft-Feynman gauge the gauge boson propagator is simplified to $-i g_{\mu\nu}/(k^2- M^2_z)$, which simplifies the whole calculation.
One can integrate out the extra scalar in the region that $\Phi$ is much heavier than the energy of the process, typically the proposed energies of ILC and TLEP: a couple of hundred GeVs. Here we do not make that assumption.

We choose to evaluate the loop diagram directly. Define the loop integral as below:
\begin{align} 
\nonumber &E_e(s,p,m_\Phi,m_h,m_z)=\int \frac{d^4k_1}{(2\pi)^4}\frac{d^4k_2}{(2\pi)^4}\\&\frac{i}{k_1^2-m_h^2}\frac{-i}{(s-k_1)^2-m_z^2}\frac{i}{k_2^2-m_\Phi^2}\frac{i}{(p-k_1)^2-m_h^2}\frac{i}{(p-k_1+k_2)^2-m_\Phi^2}\\
\nonumber &E_f(s,p,m_\Phi,m_h,m_z)=\int \frac{d^4k_1}{(2\pi)^4}\frac{d^4k_2}{(2\pi)^4}\\&\frac{i}{k_1^2-m_h^2}\frac{i}{k_2^2-m_\Phi^2}\frac{i}{(k_1-k_2)^2-m_\Phi^2}\frac{i}{(p-k_1)^2-m_h^2}\frac{i}{(p-k_1+k_2)^2-m_\Phi^2}\\
\nonumber &F_g(s,p,m_\Phi,m_h,m_z)=\int \frac{d^4k_1}{(2\pi)^4}\frac{d^4k_2}{(2\pi)^4}\\ &\frac{i}{k_1^2-m_h^2}\frac{-i}{(s-k_1)^2-m_z^2}\frac{i}{k_2^2-m_\Phi^2}\frac{i}{(k_1-k_2)^2-m_\Phi^2}\frac{i}{(p-k_1)^2-m_h^2}\frac{i}{(p-k_1+k_2)^2-m_\Phi^2}
\end{align}
p is the outgoing 4-momentum of H, and s is the center of mass energy of the process.

The way to treat higher-loop diagrams is widely studied, especially in the SM QCD calculations, and in some formal aspect of scattering amplitudes. The analytical way to do it is through the reduction of amplitudes to master integrals \cite{2011CoPhC.182..790S,2014CoPhC.185.2090S}. It is more involved than the numerical calculation.

A main challenge in computing Feynman diagrams is the evaluation of integrals over loop momenta. For multi-loop diagrams containing several mass scales we cannot realistically hope for an analytic solution; we are forced to resort to the numerical integration. The general steps are listed as follow:\\

1.Combine all propagators using a single set of Feynman parameters $x_i$. Then shift and rescale the loop momenta to reach the standard form of the denominator. \\

2.Perform the Gaussian integrals of the momenta.\\

3.Mapping Feynman parameters to the hypercube and decompose the integrals into separate ones, with IR and/or UV singularities. This step is often called \textquotedblleft Sector Decomposition \textquotedblright.\\

4.Regularize all the singularities and finish the integral. \\

The steps 3 and 4 can be very complicated and containing many singularities. However those steps are highly technical and can be done by computers. Here we evaluate our three integrals directly by using FIESTA 3.0, as functions of p and s. FIESTA stands for Feynman Integral Evaluation by a Sector decomposiTion Approach. It is based on the sector decomposition approach to the numerical evaluation of Feynman integrals originally applied by G.Heinrich and T.Binoth \cite{2004NuPhB.680..375B,2004NuPhB.693..134B}.

\subsection{Two-loop contributions in Higgsstrahlung and beyond}

The regularized differential cross section including higher order corrections in $ee \rightarrow ZH$ process is:
\beq
(\frac{d\sigma}{d\Omega})=\sum_{\sigma, \lambda}\frac{1}{4}(1+2\sigma P^-)(1-2\sigma P^+)\frac{\beta}{64\pi^2s}
\{|M_0^{\sigma,\lambda}|^2+2Re[M_0^{\sigma,\lambda}(\delta M^{\sigma, \lambda})^*]\}
\label{eezh}
\eeq
With $\sigma= \pm \half$, $\lambda=0,1,-1$ being helicities of the incoming electrons and outgoing bosons. $P^\pm$ are the degrees of polarization of the incoming fermions, such that a purely right- or left-handed electron corresponds to 1 and -1 respectively. If we assume an unpolarized beam, $P^\pm=0$. 
\be
\beta=\frac{1}{4E^2} \sqrt{[4E^2-(M_Z+M_H)^2][4E^2-(M_Z-M_H)^2]}
\ee
E is the beam energy, $S=(p_1+p_2)^2=4E^2$. $g_e^\sigma$ is the coupling of the Z-boson to left- and right-handed electrons.

Notice that at the tree level $ee \rightarrow ZH$:
\be
M_0^{\sigma,\lambda}=e^2 g_e^\sigma \frac{M_z}{s_wc_w}\frac{1}{S-M_z^2}P_1^{\sigma,\lambda}
\ee
$P_1^{\sigma,\lambda}$ contains the polarization information of Z boson, for the definition see \cite{Denner:1992bc,Denner:1991kt}.
\beq
P_1^{\sigma,\lambda}= \Big\{
\begin{tabular}{cc}
$\sqrt{2}E(\cos\theta \mp 2\sigma)$  & \text{for} $\lambda= \pm1$ \\
$\frac{4E^2+M_Z^2-M_H^2}{2M_Z} \sin\theta$ & \text{for} $\lambda=0$ 
\end{tabular}
\eeq

The $\delta M ^{\sigma,\lambda}$ here stands for the contribution of the higher loop diagrams to the invariant matrix element.

\subsubsection{Two-loop results from wavefunction renormalization}

At tree level due to the extra scalar wavefunction renormalization, the HZZ coupling becomes $ieg_{\mu \nu} M_w\frac{1}{sc^2}(1+\half \delta Z_H)$. The matrix element from this vertex correction is:
\beq
\delta M^{\sigma,\lambda}_{\text{one-loop}} = \half\delta Z_H M_0^{\sigma,\lambda} 
\eeq
The correction to the tree level diagrams due to this wave function renormalization is:
\be
\frac{\delta \sigma_{\kappa \neq 0,\text{one-loop}}}{\sigma_{SM}^{ZH}}(s,m_\Phi)= \delta Z_h
\ee

This is considered to be all the one-loop correction coming from the extra scalar, in the case that the extra scalar only couples to the Higgs sector.

For one-loop diagrams, $\delta Z_H$ correction gets into all the Higgs related couplings in forms of $\half \delta Z_H$ per Higgs line. That gives rise to:
\beq
\delta M^{\sigma,\lambda}_{\text{two-loop}} = \sum_{i}\frac{n^i_h}{2}\delta Z_H \delta M_{i,\text{one-loop}}^{\sigma,\lambda} 
\eeq
The i summing over all the SM one-loop diagrams, around 50 diagrams as calculated in \cite{Denner:1991kt}. $n_i$ here labels the number of the Higgs lines in each diagram. We are not calculating it one-by one here, and only approximate the total correction by using  Eq \ref{eezh}:

\be
\frac{\delta \sigma^{\kappa \neq 0}_{\text{two-loop, wavefunc}} }{\sigma^{ZH}_{SM}}(s,m_\Phi) \simeq a_h\delta Z_H \frac{\delta \sigma_{\text{SM,one-loop}}}{\sigma^{ZH}_{SM}}
\ee
 
 With $a_h$ being the average number of Higgs lines in each one-loop diagram. The $\delta \sigma_{SM}^{\text{one-loop}}/\sigma_{SM}$ is the one-loop SM correction comparing the tree level result. This ratio is around $5\%$ for centre of mass energy below 1TeV as calculated in \cite{Denner:1992bc,Denner:1991kt}. The equations above give rise to the two-loop correction coming from the Higgs wave function renormalization. We plot all the two loop effect combined in section IV, after the discussion  other diagrams.
 
\subsubsection{Two-loop results from diagram e,f,g}
We can write down two-loop diagram corrections to matrix elements as below by using the previous loop integrals:
\beq
&\delta M^{\sigma, \lambda}_e= i \half e^3 g^\sigma_e (\frac{m_z}{s_wc_w})^2(2\kappa i)(2\kappa v i)\frac{1}{s-m_z^2} E_e(s,p,m_\Phi,m_h,m_z)P^{\sigma, \lambda}_1\\
&\delta M^{\sigma, \lambda}_f= \frac{1}{4} e^3 g^\sigma_e /(2s_w^2c_w^2)(-2\kappa v i)^3\frac{1}{s-m_z^2}E_f(s,p,m_\Phi,m_h,m_z)P^{\sigma, \lambda}_1\\
&\delta M^{\sigma, \lambda}_g= i e^3 g^\sigma_e (\frac{m_z}{s_wc_w})^2(-2\kappa v i)^3\frac{1}{s-m_z^2} F_g(s,p,m_\Phi,m_h,m_z)P^{\sigma, \lambda}_1
\eeq
The symmetry factors of Feynman diagram are $S=2$ for Graph e, $S=4$ for Graph f, and $S=1$ for Graph g in Fig.5.

\subsubsection{Complete two-loop contributions}
So we arrive at:
\beq
\frac{\delta \sigma_{\kappa \neq 0,\text{two-loop}} }{\sigma^{Zh}_{SM}}(s,m_\Phi)=\frac{4e\kappa^2}{s_wc_w}[-m_zv iE_e+\frac{i\kappa v^3}{2m_z} E_f-4v^3\kappa m_zF_g]+\delta \sigma_{\delta_h}(s,m_\Phi)+a_h\delta Z_H \frac{\delta \sigma^{\text{one-loop}}}{\sigma^{Zh}_{SM}}
\eeq

The first three terms count the contributions from Graph e,f,g, while $\delta \sigma_{\delta_h}(s,m_\Phi)$ includes the contribution from Graph c,d, calculated from the tri-Higgs coupling corrections. And the last piece includes all the corrections to the one-loop SM diagrams. We use the result from \cite{McCullough:2013rea} to calculate the $\delta \sigma_{\delta_h}(s,m_\Phi)$. We have:

\beq
\delta \sigma_{\delta_h}= 0.013 \delta_h \kappa^2, \quad\quad
a_h\delta Z_H \frac{\sigma^{\text{one-loop}}}{\sigma^{Zh}_{SM}} \sim 0.07\delta Z_H, \quad \text{for}\quad\sqrt{s}= 250\text{ GeV}
\eeq

\beq
\delta \sigma_{\delta_h}= -0.002 \delta_h \kappa^2, \quad\quad a_h\delta Z_H \frac{\sigma^{\text{one-loop}}}{\sigma^{Zh}_{SM}} \sim 0.05\delta Z_H,\quad \text{for}\quad\sqrt{s}= 500 \text{ GeV}
\eeq

Now choosing the the center of mass energy $\sqrt{s}=250, 500 \text{ GeV}$ and combining the results, we can plot the dependence of the correction to the total cross section on the mass of extra scalar $\Phi$. Notice that this is the two-loop contribution.  The one-loop result is only through wave-function renormalization:  $\delta \sigma_{\text{wavefunction}}/ \sigma_{SM} = \delta Z_h$ as plotted in Fig.3. The two-loop contribution comes in different sign in most of the range above comparing to the one-loop contribution, and comparable by size.

\begin{figure}[t]
\subfigure[]{
\includegraphics[width=8cm,natwidth=610,natheight=632]{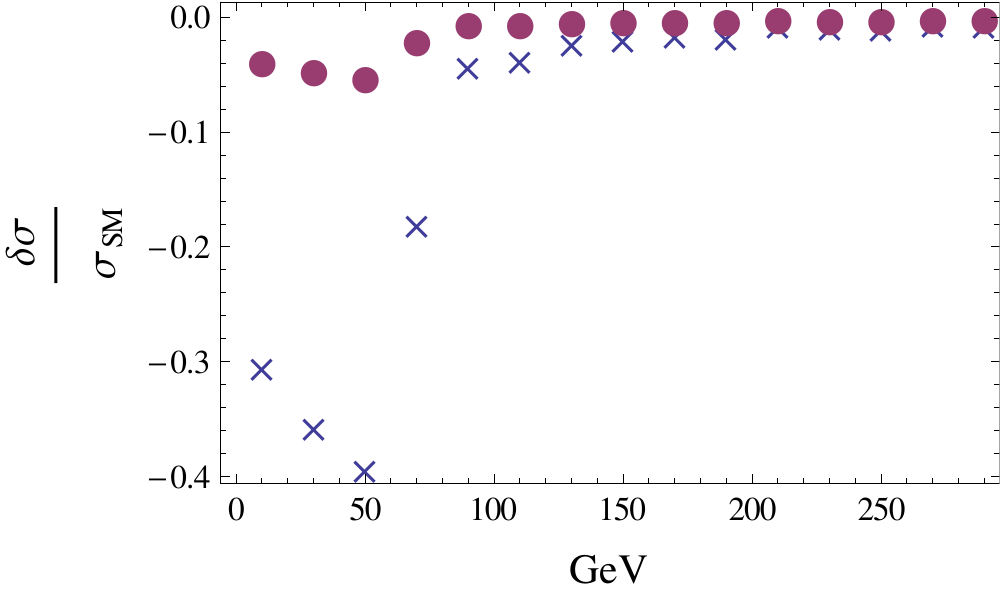}}
\subfigure[]{
\includegraphics[width=8cm,natwidth=610,natheight=632]{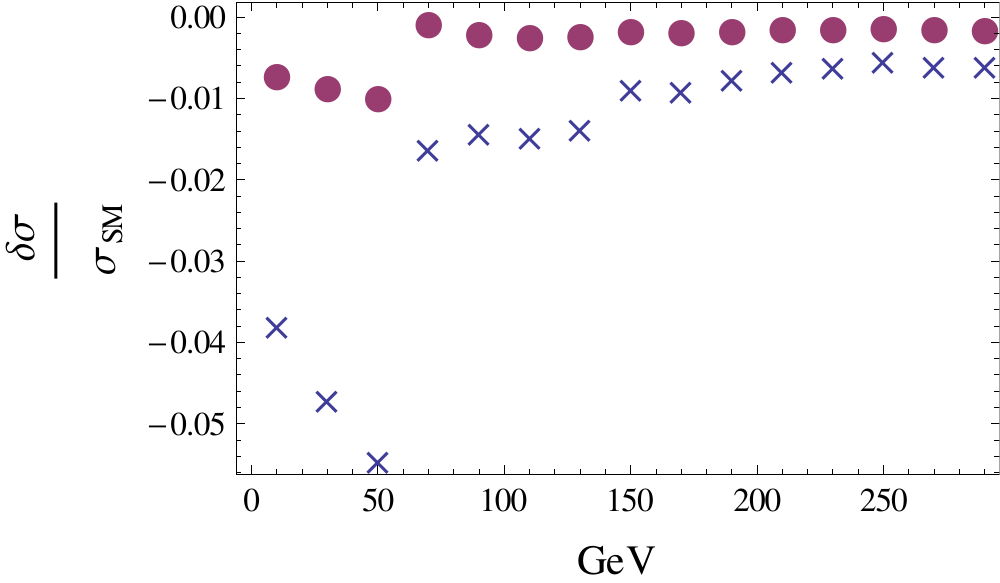}
}
\caption{ $\delta \sigma_{\kappa \neq 0}/ \sigma^{Zh}_{SM}$ versus $m_\Phi$ with $\sqrt{s}=250$ GeV (a) and $\sqrt{s}=500$ GeV (b), assuming $\kappa=2$ for crosses, $\kappa=1$ for round dots.}
\label{fig:tri}
\end{figure}

\section{The connection to Electroweak Baryogenesis}

There has been extensive studies about the constraints on a first-order electroweak phase transition with a single scalar extension to the SM \cite{Curtin:2014jma,Katz:2014bha}. This is sometimes called \textquotedblleft nightmare scenario", for the reason that it is hard to be probed  and only contains several slim parameter regions where the strong electroweak phase transition can be achieved. The precision measurement of the Higgsstrahlung cross section can potentially reduce allowed parameter space according to the two-loop result presented in this paper, especially for a gauge singlet scalar. Our result here shows that to correctly constrain those parameters through Higgsstrahlung, one needs to include the two-loop effect, for the scalar mass being smaller than a couple of hundred GeVs. One can see \cite{Curtin:2014jma} for more discussion on the parameter space of the EWBG. This parameter region will be further constrained by a virtual $h^* \rightarrow \Phi\Phi$, tri-Higgs couplings and the Higgsstrahlung measurements in next generation colliders.

The loop correction in tri-Higgs coupling provides an interesting new source for CP violation needed in electroweak baryogenesis. This effect can come from a single Higgs doublet and an extra real scalar rather than the standard two-Higgs doublet model \cite{1993ARNPS..43...27C}. Whether this new CP violation source is enough to achieve the EWBG remains unknown.

The two-loop thermal correction to Higgs potential due to the extra scalar is beyond the scope of this paper. However we expect it be sizable as well and place non-trivial bounds on the EWPT. 

\section{Conclusions}
In this paper, we calculate the two-loop contribution to the Higgsstrahlung process for a particular BSM scenario. We have included one more real singlet scalar to the SM. This extra scalar couples to the SM Higgs with $Z_2$ symmetry. We divide all the two-loop contributions into three parts: tri-Higgs coupling correction, wavefunction renormalization and the others. The two-loop correction partly cancels out the one-loop correction especially when the mass of the extra scalar is below the Higgs mass. Next generation lepton colliders will measure the Higgsstrahlung cross section within the precision of the two-loop effect. The loop effect from this real scalar can also induce a new CP phase to the tri-Higgs coupling when the mass of this real scalar is smaller than half of center of mass energy.

Although these loop calculations are valid even if this real scalar is charged under electroweak $SU(2)$ or color, there are more loop diagrams involved in Higgsstrahlung process for charged extra scalars. The contributions from those diagrams are not necessarily larger than the ones present in this paper. Furthermore colored extra scalar will get more constraints from Higgs production cross sections in gluon fusion. Electroweakly charged extra scalar can modify $BR(h\rightarrow \gamma\gamma)$ which is potentially a powerful observable. The Higgs couplings to all SM particles have already been modified regardless of the quantum number of the extra scalar, as we have shown in section II. In the \textquotedblleft nightmare scenario\textquotedblright, when the scalar singlet is considered to be a possible dark matter candidate, we have very limited observables to probe it rather than the Higgsstrahlung cross section.

\begin{figure}[t]
\begin{center}
\subfigure[]{
\includegraphics[width=5cm,natwidth=610,natheight=632]{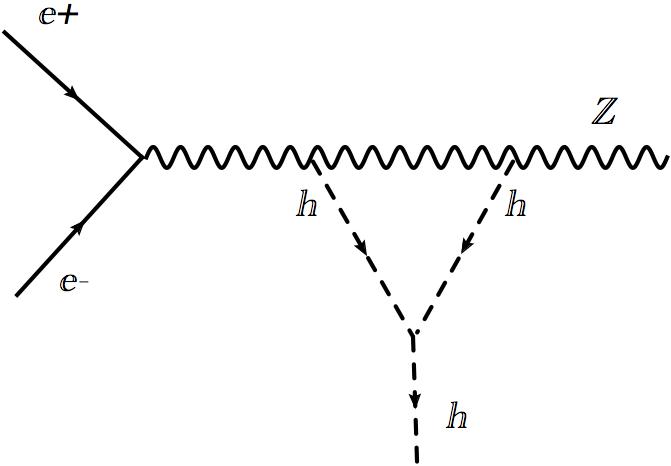}}
\quad\quad\quad
\subfigure[]{
\includegraphics[width=5cm,natwidth=610,natheight=632]{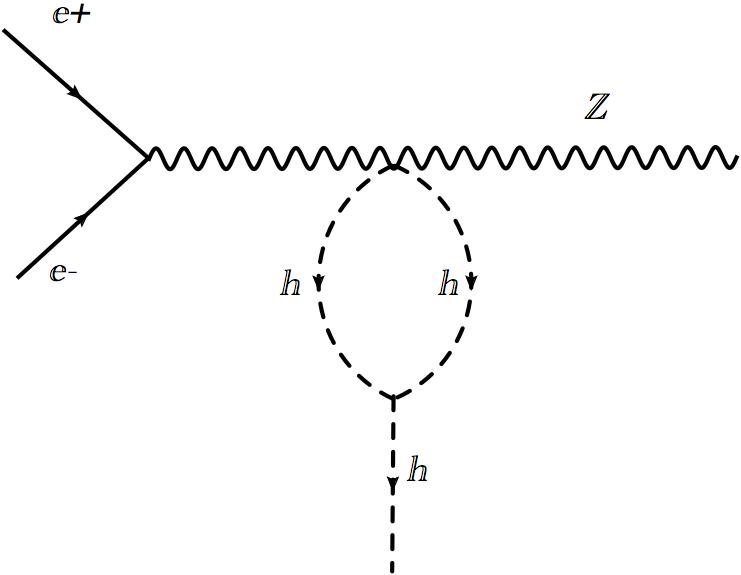}
}
\quad
\subfigure[]{
\includegraphics[width=5cm,natwidth=610,natheight=632]{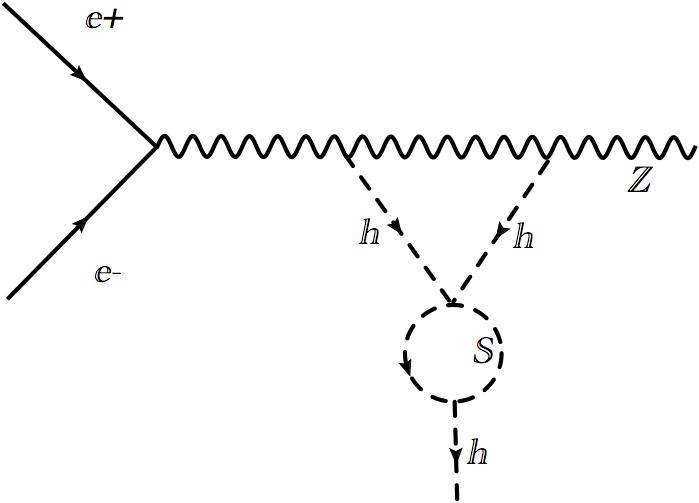}
}\subfigure[]{
\includegraphics[width=5cm,natwidth=610,natheight=632]{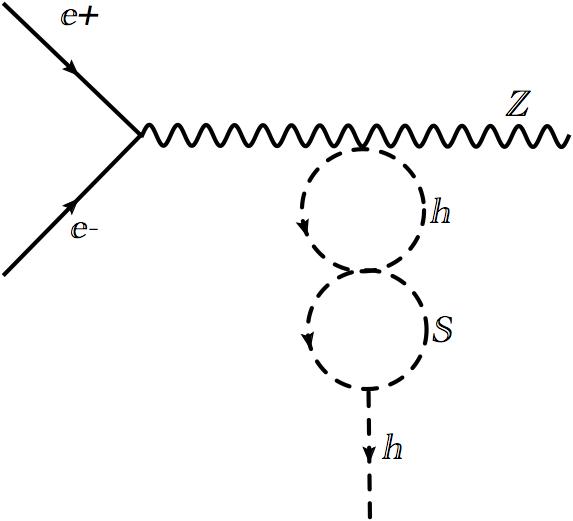}
}
\subfigure[]{
\includegraphics[width=5cm,natwidth=610,natheight=632]{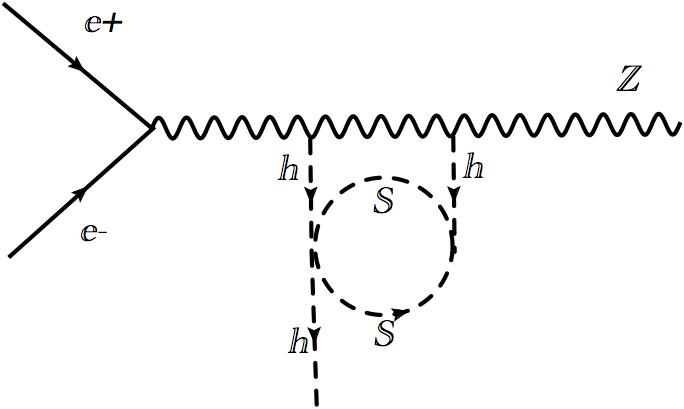}
}
\subfigure[]{
\includegraphics[width=5cm,natwidth=610,natheight=632]{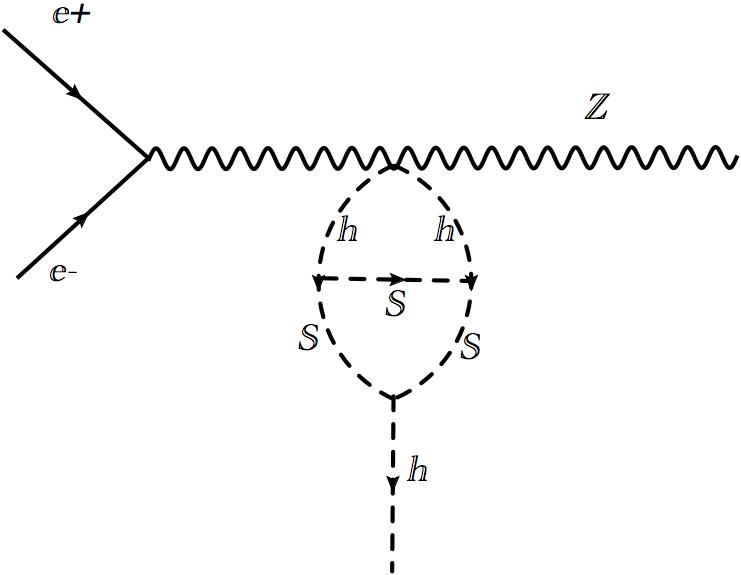}
}
\subfigure[]{
\includegraphics[width=5cm,natwidth=610,natheight=632]{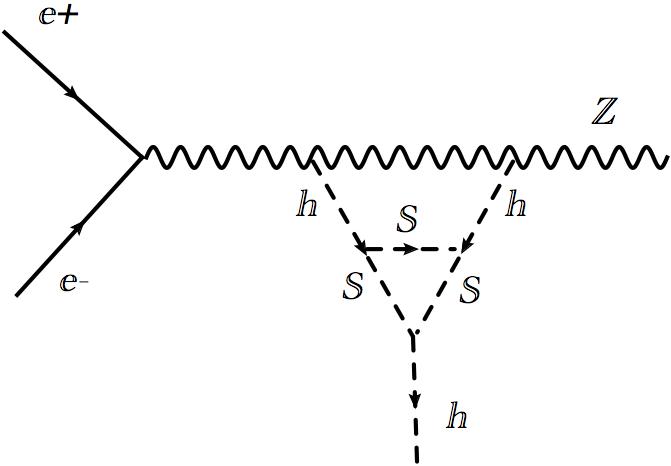}
}
\caption{The one and two loop Feynman diagram contributions to Higgsstrahlung process with a gauge singlet scalar.}
\label{fig::twoloop}
\end{center}
\end{figure}

\begin{acknowledgments}
 I would like to thank Lian-tao Wang and Tao Liu for the early involvement of this work and Hua-xing Zhu for introducing the FIESTA program. I thank Jan Hajer for commenting on the draft. This work was supported by the CRF Grants of the Government of the Hong Kong SAR under HUKST4/CRF/13G. 
 \end{acknowledgments}

\bibliography{eezhbib.bib}

\end{document}